# Deep Learning Based Multi-Label Text Classification of UNGA Resolutions

Francesco Sovrano [b]

Monica Palmirani [a]

Fabio Vitali [b]

[a] CIRSFID, University of Bologna
[b] DISI, University of Bologna

## ABSTRACT

The main goal of this research is to produce a useful software for United Nations (UN), that could help to speed up the process of qualifying the UN documents following the Sustainable Development Goals (SDGs) in order to monitor the progresses at the world level to fight poverty, discrimination, climate changes. In fact human labeling of UN documents would be a daunting task given the size of the impacted corpus. Thus, automatic labeling must be adopted at least as a first step of a multi-phase process to reduce the overall effort of cataloguing and classifying. Deep Learning (DL) is nowadays one of the most powerful tools for state-of-the-art (SOTA) AI for this task, but very often it comes with the cost of an expensive and error-prone preparation of a training-set. In the case of multi-label text classification of domain-specific text it seems that we cannot effectively adopt DL without a big-enough domain-specific training-set. In this paper, we show that this is not always true. In fact we propose a novel method that is able, through statistics like TF-IDF, to exploit pre-trained SOTA DL models (such as the Universal Sentence Encoder) without any need for traditional transfer learning or any other expensive training procedure. We show the effectiveness of our method in a legal context, by classifying UN Resolutions according to their most related SDGs.

## 1. INTRODUCTION

In the Thirty-Third Session of the High Level Committee on Management (HLCM) held in Budapest, 30-31 March 2017[2], the United Nations adopted the Akoma Ntoso XML standard for the United Nations System (AKN4UN[3]) as well as the United Nations System Document Ontology (UNDO) to provide a formal representation of the fundamental entities of UN documents and of their relationships.

HLCM adopted the AKN4UN Guidelines for the markup of UN normative and parliamentary documents, and UNDO ontology as the main reference model for the implementation of UN-SIF, the United Nations Semantic Interoperability Framework, in order to identify the structural parts and the semantic aspects of the sentences according to the specific goals of each UN Agency or Department. One common task is to qualify the UN documents following the Sustainable Development Goals (SDG)[4] in order to

---

monitor the progresses at the world level to fight poverty, discrimination, climatic changes.

In 2015, 17 Sustainable Development Goals (SDGs) and 169 targets were adopted by the world leaders and in 2016 this list officially came into force. Those goals define the Agenda for Sustainable Development till 2030 and the intention is to universally apply them to eradicate poverty, fight inequalities, tackle climatic change, support inclusion. Progresses are monitored using 232 unique indicators[5] and open data are used for such information. One of the most important task in the SDG approach is to detect the actions connectable with the targets that allow to better measure the effectiveness of the word-wide policy. The actions implementing the SDG are many and the information related to their success are collected using different sources.

An interesting and worthy task within this framework is therefore the classification of UN documents, as well as other kinds of documents, according to the above-mentioned SDG goals and targets, so as to detect trends and indicators and to produce open datasets [1] useful for statistics and predictors, and consequently to better inform political strategies of the UN and participating countries to reach such goals. The SDGs are mentioned inside of the text in explicit manner using a regular citation pattern (e.g., SDG, target 11.2, see Fig. 1). However several sentences are indirectly referred to SDGs using keywords in target definitions (see the Fig. 1 the yellow fragment of text). Of course, human labelling of UN documents, according to the goals and targets, would be a daunting task given the size of the impacted document corpus. Automatic labelling, therefore, must be adopted at least as a first step of a multi-phase process to reduce the overall effort of cataloguing and classifying. Deep Learning approaches seem the most appropriate tools for multi-label classification of text documents, but the requirements for a reasonably large training set of pre-labelled documents is not necessarily within reach for this type of projects.

In this paper we discuss a new method, able to provide effective multi-label annotation and a decent accuracy. This new method uses Deep Learning based technology without a traditional training set, by exploiting statistical methods such as TF-IDF.

Using a technique called Text Similarity Approach (TSA), we were able to establish a working approach, annotating UN resolutions for a dev-set of around 30 manually annotated paragraphs, used to find the optimal value for the discriminative threshold used to identify whether a SDG is related or not to a paragraph. Our dev-set is much smaller than the training-set used in other approaches, such as [2] that use around 16000 annotated documents for training a multi-label classification model on the "20 Newsgroups" dataset.

In the following, we discuss the methodology we have used, and provide some initial results about the accuracy and sophistication of our annotator. In particular, in Section 2 we illustrate the problem space and the methodology. In Section 3 we introduce the required background information to understand our work, in Section 4 we discuss the architecture of our system. In Section 5 we provide some implementation details and in Section 6 and 7 we show the experimental results and we discuss the results we got.

---

In section 8 we compare our proposed technique with related work. Finally, in section 9 we provide some additional concluding remarks and suggest pathways for future works.

## 2. APPROACH

The UN has defined 17 different SDGs (e.g., no poverty, no hunger, good health and well-being, quality education, etc.). Every SDG is about a different topic and has several targets. Thus, we have to classify the content of every resolution according to 18 classes (17 classes for the SDGs, and 1 class for everything else).

UN Resolutions are the texts of the formal expressions of the opinion or will of United Nations organs. All the UN General Assembly Resolutions are publicly available on the UN website [3]. A UN resolution has a regular structure composed of a preface with the title and the identification information (date and number), a preamble with justificatory and introductory paragraphs, a full body with the actual norms, and conclusions part. The multi-label text classification of a UN Resolution can be performed at different granularities of the document, for example at the document level or approaching each paragraph separately. We decided to work at the paragraph level, because:

1) A paragraph is smaller than the whole document and thus, intuitively, it is easier to classify correctly.

2) We have 609 UN Resolutions with a grand-total of 26784 paragraphs. Intuitively, it is harder to reliably associate a multi-label classification to 609 texts, while it becomes more appropriate with 26784 texts.

**Resolution adopted by the General Assembly on 23 December 2015 70/248. Special subjects relating to the proposed programme budget for the biennium 2016–2017**

*Recalling* that the Sustainable Development Goals and targets are integrated and indivisible and balance the three dimensions of sustainable development, and acknowledging the importance of reaching the road safety-related targets, such as target 3.6, which aims to halve, by 2020, the number of global deaths and injuries from road traffic accidents, and target 11.2, which aims to provide, by 2030, access to safe, affordable, accessible and sustainable transport systems for all, improving road safety, notably by expanding public transport, with special attention to the needs of those in vulnerable situations, women, children, persons with disabilities and older

**Figure 1: Explicit and Implicit Sustainable Development Goals**

A UN Resolution is a structured document defined by its informative content and presented according to a pre-defined template. In order to effectively separate the presentation from the informative content, we automatically convert every UN Resolution from Microsoft Word format into Akoma Ntoso [4] using parsers, regular expressions and patterns.

In this particular setting, every SDG states a well-defined and different concept, using sub-topics called targets. In Akoma Ntoso any classification is made using the particular metadata block <keyword> that is capable of assigning labels to document fragments:

```xml
<classification source="#cirsfidUnibo">
  <keyword eId="keyword_5_5_2"
value="goal_5_5.2" href="#para_3 #para_7"
showAs="SDG 5_5_2" refersTo="#concept_s-
dg_5_5_2" dictionary="SDGIO"/>
```

This fragment of XML associates the keyword of goal_5_5_2 to the paragraphs 3 and 7 of the document, and in meantime it provides the label used by the end-user during the searching (SDG 5_5_2) and the source of the dictionary (SDGIO).

Each keyword is also connected with a Top Level Class (TLC) Concept using the @refersTo attribute. The TLCs are basic classes of the non-ontology of Akoma Ntoso [29][30] that permits to connect each fragment of the document to the real ontology (in this case the SDGIO[6]).

```xml
<TLCConcept eId="concept_sdg_5_5_2" href="/
akn/ontology/concepts/un/sdg/sdgio/goal_5_5_2"
showAs="SDG 5.5.2"/>
```

In the proprietary block of metadata we store the confidence of the classification with also a recap for each paragraph concerning its multi-level classification.

```xml
<proprietary source="#cirsfidUnibo">
    <akn4un:source href="#para_3">
        <akn4un:sdgTarget value="goal_5_5_2"
confidence="1.6334762573242188" name="SDGIO"/>
        </akn4un:source>
    <akn4un:source href="#para_7">
        <akn4un:sdgTarget value="goal_5_5_2"
confidence="1.9220209121704102" name="SDGIO"/>
        </akn4un:source>
</proprietary>
```

Classifying a resolution paragraph according to its most related SDGs is equivalent to understand whether the concepts expressed in the paragraph are similar enough to one or more SDGs. In other words, our multi-label text classification problem is also a sort of multi-concept recognition problem.

There are at least two approaches to tackle this peculiar classification problem: a Classical Approach, and a Text Similarity Approach. In the Classical Approach we train an AI to minimize classification errors (e.g., the cross-entropy of some learned categorical distribution), given a big-enough annotated training set. Training from scratch usually requires much more annotated training data. For instance, [2] use around 16500 annotated documents to train a deep learning based model for a similarly sized (20 classes) multi-label classification on the "20 Newsgroups" dataset. Transfer learning can help to reduce the amount of annotated data required for good results [5], but can be an error-prone and slow procedure that might introduce some unwanted bias.

In the Text Similarity Approach, on the other hand, we compute how similar is a source text (a paragraph of a UN Resolution) to a target text (the definition of a SDG) by using some vectorial representation of these texts, and if the similarity is above a certain discriminative threshold then we can say that the source expresses concepts similar to those expressed by the target. These vectorial representations may be obtained by applying transfer learning in an unsupervised manner to some pre-trained model. But unsupervised transfer learning does not guarantee us that, during transfer, the algorithm is able to encode (in the embedding) enough information about how to classify a sentence according to its relatedness to some class. How, then, can we effectively exploit deep learning models without traditional transfer learning nor big annotated training sets?

We think we can adopt the aforementioned Text Similarity Approach without any need for (supervised or unsupervised) traditional transfer learning, even if the pre-trained model has been trained on tasks and datasets unrelated to the UN Resolutions domain. The new model we propose is an hybrid model and it is based on the combination of statistical models, as TF-IDF [6], with pre-trained state-of-the-art deep learning models, as the Universal Sentence Encoder [5]. The statistical models are used to extract domain- specific information and are built using only the definitions of the SDGs. While the deep learning based models are pre- trained in an unsupervised manner on very generic datasets and used to extract only generic and language-dependent information.

## 3. BACKGROUND

At a high level, our proposed algorithm works by representing both the paragraphs and the SDG definitions through numerical encodings. Several techniques exist for learning numerical representations of words from their occurrence information, we can group these numerical representations into two main categories: Scalars (eg. TF-IDF), and Vectorial (eg. the word embeddings).

### 3.1 Word Embeddings

The term "word embedding" has been originally coined by Bengio et al. [7]. Word embedding is a type of mapping that allows words with similar meaning to have similar representations. The basic idea behind word embedding (and distributional semantics) can be summed up in the so-called distributional hypothesis [8]. Word2Vec [9], GloVe [10] and fastText [11] are unsupervised learning algorithms for word embedding, based on Artificial Neural Networks. All the aforementioned word embedding algorithms consist in an Artificial Neural Network (ANN) usually trained by mean of Stochastic Gradient Descent, intuitively with the goal of optimally predicting a word given its context or vice versa.

An important aspect of these embeddings is the ability to solve word analogies in the form "A is to B what C is to D", by using simple arithmetic. For example, in Word2Vec, we might see that the following word embeddings equations are valid: "Paris - France + Germany= Berlin", "King - Man + Woman = Queen". Thus, the similarity between these embeddings is said to be paradigmatic and it is usually measured through cosine similarity.

---

## 3.2 Document Embedding

Document (or sentence) embedding is somehow related to word embedding, but it is a different task because the granularity of the input of the embedder shifts from words to documents. Some famous document embedding techniques are: Bag of Words [12] (BoW), Term Frequency - Inverse Document Frequency [6] (TF-IDF), Average Word Embedding [9] (AWE), Universal Sentence Encoder [5] (USE).

### 3.2.1 Term Frequency - Inverse Document Frequency

TF-IDF [6] is both a word and a document encoding technique. Documents encodings are based on BoW [12]. In BoW, documents are described by word occurrences while completely ignoring the relative position information of the words in the document. BoW tokenizes the documents, counts the occurrences of tokens, and returns them as a sparse matrix.

The TF-IDF is the product of two statistics: Term Frequency (TF) and Inverse Document Frequency (IDF). TF is basically the output of a BoW model. For a specific document, it determines how important a word is by looking at how frequently it appears in the document.

On the other hand the IDF statistic is based on the idea that important document words (also called signature words) appear frequently within the same document but rarely within different documents. Thus, the frequency of a signature word must be low among different documents, in other words the Inverse Document Frequency must be high. BoW and TF-IDF can produce document embeddings.

The similarity between TF-IDF document embeddings is said to be more topical (topic-related) or syntagmatic [13] and it is usually measured through cosine similarity.

### 3.2.2 Average Word Embedding

A naive approach to build document embeddings might be averaging the word embeddings of a document, this is called Average Word Embedding (AWE). One of the disadvantages of this document embedding technique is that it is not sensible to words ordering.

Intuitively, averaging the word embeddings of a document is not the only way we can combine word embeddings in order to produce a document embedding. In fact many other approaches exist for combining the word embeddings of a document, some of them involve TF-IDF [15,16].

### 3.2.3 Universal Sentence Encoder

A more sophisticated approach for document embedding might be the Universal Sentence Encoder (USE). Two variants of the USE have been proposed in [5]: the Transformer-based and the DAN-based. The Transformer-based takes its name from the homonymous [17] Deep Neural Network, and has higher accuracy than the other variant, but it has quadratic complexity with respect to the input size. The DAN-based takes its name from the Deep Averaging Network [5], and has linear complexity, but apparently lower accuracy. Differently from AWE, USE learns to embed the whole sentence directly in an end-to-end manner, providing state-of-the-art results.

## 4. DEEP LEARNING BASED MULTI-LABEL TEXT CLASSIFICATION

We have to identify whether a paragraph of an official English resolution of the United Nations (UN) is related to one or more Sustainable Development Goals. Furthermore, every goal (SDG) may have different targets that may change in the near future (eg. some of the targets have a short- or mid-term deadline: 2020, 2030). Thus, we need a Natural Language technique that should respect at least the following requirements:

1) The algorithm should be able to decide whether a given paragraph is related or not to a SDG.
2) The (learning) algorithm should require almost no annotated training set for properly working, and should allow us to easily change the SDG definitions without incurring significantly slower or more error-prone pre-processing (eg. a slow model-training phase).

As described in Section 2, we can meet the first aforementioned requirement by computing how similar is a source text (a paragraph of a UN Resolution) to a target text (the definition of a SDG), and if the similarity is above a certain threshold then we can say that the source expresses concepts similar to those expressed by the target.

In Section 3 we have seen that many models exist for document embedding, but we are going to study only some of them: TF-IDF, Average GloVe, and the Universal Sentence Encoder.

The first model (TF-IDF) is probably the fastest to build/train, especially because it does not require any labelled dataset nor hyper-parameters tuning. While the other models are slower to train and they usually depend on a lot of hyper- parameters.

Models for word or document embedding, pre-trained on very big and generic datasets, are available on the web, but these pre-trained models are usually not optimized for domain- specific tasks.

Every SDG has an official English description publicly available at [18]. But these descriptions alone seem to be not enough for properly training a ANN-based model from scratch, nor for traditional transfer learning.

Thus we designed a new ensemble method that effectively combines: generic (non domain-specific) document similarities obtained through pre-trained models (GloVe and USE), with domain-specific document similarities obtained through TF- IDF. This way we avoid any complicated and error-prone learning phase for building ad hoc document embedding models, thus allowing us to easily tackle also requirement 2 without losing the benefits of deep learning based techniques.

In other words, our solution tries to exploit the best from two distinct techniques. TF-IDF is used to model domain-specific information (by building the TF-IDF model on the UN Resolutions), but it is a shallow learning technique and it lacks of semantic expressiveness when compared to techniques such as GloVe or USE. While the pre-trained models are used to model only generic information (eg. semantic relationship among non domain-specific words).

The pre-trained models we are going to use are: a GloVe model from Spacy [19] and pre-trained on data from Common Crawl [20], and the Universal Sentence Encoder (USE) model for document embedding coming from TensorFlow Hub [5]. These models have been trained on data unrelated to the UN Resolutions, thus the resulting embeddings tend to lose information when used in specific domains such as the UN Resolutions. This is why we use TF-IDF for handling domain- specific information, as introduced is Section 2.

More in detail, let A (the query; a paragraph) and B (a corpus document representing a SDG) be two distinct documents, we want to compute the similarity between A and B. In order to do that, we combine the cosine similarity of the TF-IDF embeddings of A and B with the cosine similarity of the USE embeddings weighted by the cosine similarity of the Average GloVe embeddings.

The TF-IDF document similarity is a sort of topical similarity extracted by populating the vectors with information on "which text regions the linguistic items occur in". While both the Average GloVe and USE similarity are a sort of paradigmatic similarity extracted by populating the vectors
with information on "which other linguistic items the items co-occur with".

In other words, the idea behind this ensemble is to combine the unique and different properties of the aforementioned similarities, in order to get a new paradigmatic similarity potentially able to express topical similarity in a domain on which the pre-trained models have not been trained on.

## 5. IMPLEMENTATION DETAILS

The pipeline of our algorithm is defined by the following 5 steps: corpus pre-processing, TF-IDF model building, query pre-processing (same as corpus pre-processing, but for queries), query similarity computation, and query classification.

### 5.1 Corpus and Query Pre-Processing

Corpus and Query Pre-Processing can be resumed by the following instructions:

1) Replace upper-cases with lower-cases.
2) Replace every occurrence of "sustainable development goal" with "sdg".
3) Replace every occurrence of "sdg" followed by a cardinal number in [1,17] or preceded by an ordinal number in [1, 17], with the concatenation of "sdg" and that number (eg. "second sdg" becomes "sdg2").
4) Perform tokenization and lemmatization
5) Perform stemming on lemmas, by using the Snowball algorithm [23].
6) Remove stop-words, as defined by Spacy [19], and punctuation.

We have empirically observed that stemming helps TF-IDF in achieving greater generalization and better results in SDG classification.

We decided to consider the words "Sustainable Development Goal" as a unique token and furthermore to give a unique identifier to every SDG ("SDG1" stands for the first SDG, and so on), this is the reason behind the replacements described before. We took this decision in order to better classify all those SDGs explicitly mentioned through their unique identifier.

### 5.2 TF-IDF Model Building

Before building a TF-IDF model we need to define a corpus of documents used to extract the signature words. We build the TF-IDF model only once, before any query, by populating a fixed dictionary of all the possible words in the corpus.

In this setting, the corpus is defined by 34 different documents, two for every SDG. Every SDG is represented in the corpus by a class document and a bias document.

A class document is the description of the SDG available at [18] concatenated with the unique identifier (ID) of the SDG itself. This concatenation is performed in order to be able to correctly classify queries containing the unique IDs.

The unique ID of a SDG is an important marker for SDG classification, but even by applying the aforementioned pre- processing tricks and by using the bias documents the TF- IDF model is not able to understand the importance of the unique identifier because these IDs are single unique tokens (thus with very low term frequency in documents with more than a couple of tokens). For this reason we need the bias documents in order to add inductive bias toward these IDs. A bias document is simply a document containing only the SDG's ID.

### 5.3 Query Similarity Computation

Our corpus for the SDG classification is made of N = 34 documents. After we built the TF-IDF model we can compute a query similarity as follows:

1) Get the BoW of a query Q using the (fixed) corpus dictionary, and compute its TF-IDF vector.
2) Compute the TF-IDF cosine similarity F between the query vector and the vector of every document in the original corpus. The result should be a vector F of N real numbers in [0, 1], each one representing the similarity of the query and a document in the corpus.
3) Compute the GloVe AWE cosine similarity G between the whole corpus and the query. The result should be a vector of N real numbers in [−1, 1].
4) Compute the cosine similarity U between the whole corpus and the query embedding obtained by the DAN-based Universal Sentence Encoder [5].
5) Compute R: the squared average of G. R should be a measure of how much Q is relevant to the corpus topic (the topic of SDGs).
6) Compute $C=(F+U)\cdot R$. Where C is called combined similarity and U is called semantic shift, while R is called paradigmatic topic weight.

The intuitive idea behind using the semantic shift and the paradigmatic topic weight is that the TF-IDF similarity F is high for a query Q and a document D when the query words and the document words are similar, but F is more a syntagmatic similarity and thus may be lower (or even 0) when Q contains words in the synsets of D. Thus, in order to address the aforementioned synset-words problem we sum F with a paradigmatic similarity G before scaling it by R. We scale (F + G) by R in order to give significantly more similarity to the queries paradigmatically more related to the corpus topics.

## 5.4 Query Classification

Now that we have the combined similarity, we can use it in order to perform SDG classification. A query can be classified as related to one or more SDGs or not. Thus, we have to understand when a query is not related to any SDG. In order to do this, we have to choose a similarity threshold T.

Empirically we observe that the bigger is the query, the smaller tends to be the value of C. Thus we hypothesize that T is a function of the size of the query, for this reason in order to perform SDG classification using C we perform the following steps:

1) We compute $W=C\cdot(1+L)$, where L is the binary logarithm of the number of tokens in the query. This step is called log-length scaling.

2) For every class (SDG) we sum the weighted similarity of the bias document to the weighted similarity of the class document, thus obtaining the biased similarity B.

3) Let M be the average of B, we compute $B=B-M$ in order to centre the biased similarity vector B. The goal of centering B is to give more focus on the variance of the query-corpus similarity.

4) We sort the class documents D by descending biased similarity B, and we get the index of all the class documents V having $B > T$.

5) If the set of V is empty, then the query Q is said to be not related to any class. Otherwise we have the ranking of the most related classes to Q.

The intuitive idea behind the scaling of C by L is that the bigger is the query Q, the (smoothly) lower is C. We sum 1 to L before scaling because otherwise queries having length 1 would have W equal to 0. Queries having length 1 may be reasonable (eg. a query containing only a SDG's ID).

## 6. VALIDATION AND EVALUATION

We annotated 5 different datasets:

- **Dev-Set**: used to tune the algorithm during development. This dataset has 36 annotated elements by A. These elements do not appear in the other datasets.
- **Test-Set A**: it contains 121 paragraphs manually annotated by A.
- **Test-Set B**: it contains 105 paragraphs manually annotated by B. This set shares 50 paragraphs with set A with possibly different annotations.
- **Test-set CB**: it contains 995 paragraphs annotated by CB.
- **Test-set CL**: it contains 995 paragraphs annotated by CL. The paragraphs in this set are the same of set CB, but with possibly different annotations.

Test-sets Dev/A/B and CB/CL have different sizes and have been annotated using different approaches.

Test-sets CB and CL are the biggest datasets and have been annotated by two different (and usually expensive) legal experts.

These legal experts annotated the dataset confirming (or not) each label provided by a multi-label classification algorithm (our CDM-Transformer with threshold T = 0.6), thus without adding any new label to the set of labels provided by the algorithm. This particular way of annotating it was very fast, but with the cost of many false negatives in the annotations.

Test-sets Dev, A and B are the smallest and have been manually annotated without the support of any classification algorithm. Annotating these dataset was very expensive in terms of time, but they actually have much less false negatives in the annotations.

All the aforementioned datasets are imbalanced, in fact most of the annotated labels are of type 0 (no SDG), or 16 ("Promote just, peaceful and inclusive societies") or 17 ("Revitalize the global partnership for sustainable development"). This unbalancedness problem is much more evident in test-sets CB and CL, as shown in Table 1.

**Table 1: Class distribution comparison in datasets.**

| SET | No SDG | SDG 16 | SDG 17 | Remaining SDG |
|-----|--------|--------|--------|---------------|
| Dev | 28,8% | 11,1% | 13,3% | 46,8% |
| A | 28,5% | 25,7% | 8,5% | 37,3% |
| B | 42,8% | 19,6% | 1,7% | 35,9% |
| CB | 33,3% | 23,6% | 30,9% | 12,2% |
| CL | 31,8% | 22,2% | 34,9% | 11,1% |

## 6.1 Evaluation Metrics

Depending on the target problem, the evaluation measures for multi-label classification can be grouped into 3 main categories (according to [24]): evaluating partitions (eg. F1-Measure, Precision, Recall, Accuracy, etc.), evaluating ranking (eg. Coverage error, Ranking loss, Label ranking average precision, etc..), using label hierarchy.

We are interested in the first two categories. The evident unbalancedness of the dataset makes the evaluation of the algorithm much harder to accomplish, for this reason we decided to adopt: the Label Ranking Average Precision, the weighted F1, the Best-Ranked (BR) Accuracy and the BR weighted F1.

Best-Ranked statistics are the statistics of the best ranked label in the intersection of true labels with the predicted labels. If the aforementioned intersection is empty, then a random true label is taken. BR statistics seem reasonable due to the fact that the average labels per target (paragraph) is very low: between 1 and 1.5 depending on the test-set. Here, the BR Accuracy is equivalent to the BR micro F1.

The weighted F1 is a variant of the macro F1, weighted by support (the number of true instances for each label) in order to address unbalancedness.

We computed all the aforementioned statistics by using the metrics provided by Scikit Learn [25].

## 6.2 Experiments and Results

We want to perform multi-label text classification of UN Resolutions through deep learning, but without any big-enough annotated training set.

Our baseline algorithms for solving the aforementioned problem are classification algorithms based on USE or AWE, pre-trained on non domain-specific datasets. With our experiments we show that we can improve over the baselines without any need for traditional transfer learning or re-training the underlying neural network. We do it by adding domain-specific information extracted through TF-IDF from the UN Resolutions.

In this section we discuss an ablative analysis of our hybrid model (CDM). In order to do this, we compare our algorithm with the baselines, while changing the value of the threshold T (see Section 5.4 for details on T). In figure 2, 3, 4, 5 and 6 we show the statistics obtained respectively on the dev-set and test-sets A, B, CB and CL.

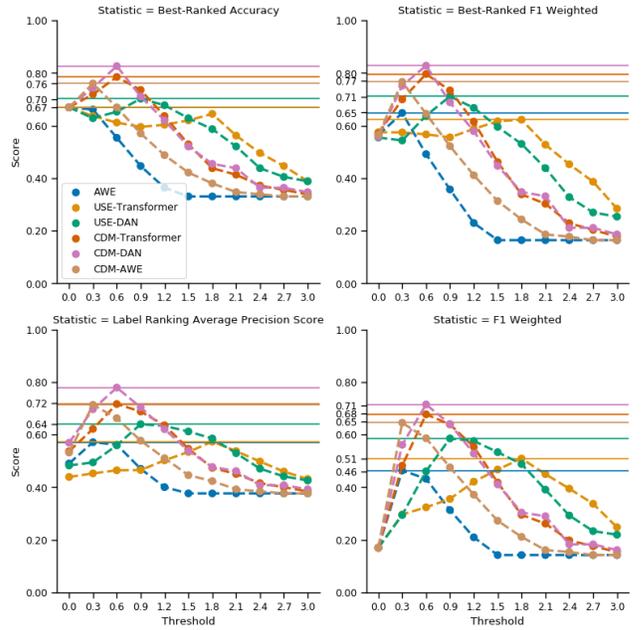

**Figure 3: Test-Set A - Score comparison of the statistics of different algorithms when changing the threshold.**

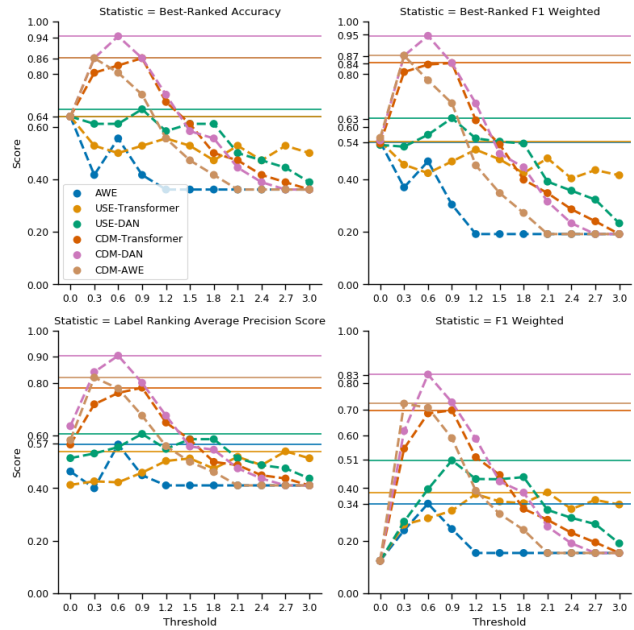

**Figure 2: Dev-Set - Score comparison of the statistics of different algorithms when changing the threshold.**

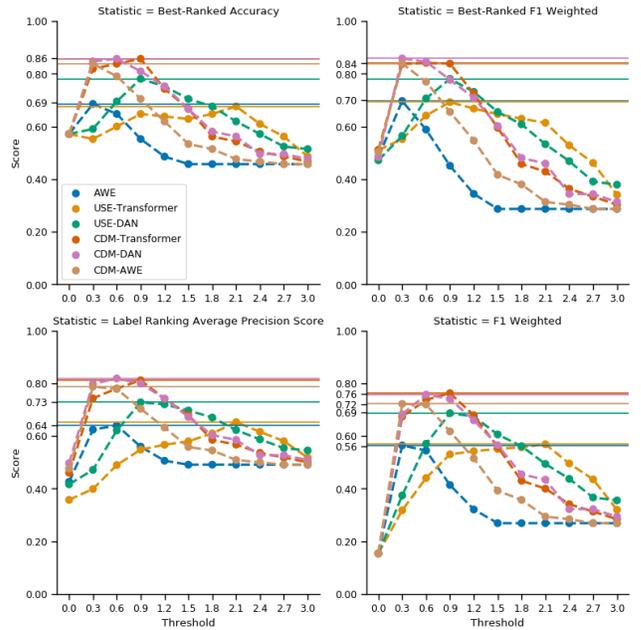

**Figure 4: Test-Set B - Score comparison of the statistics of different algorithms when changing the threshold.**

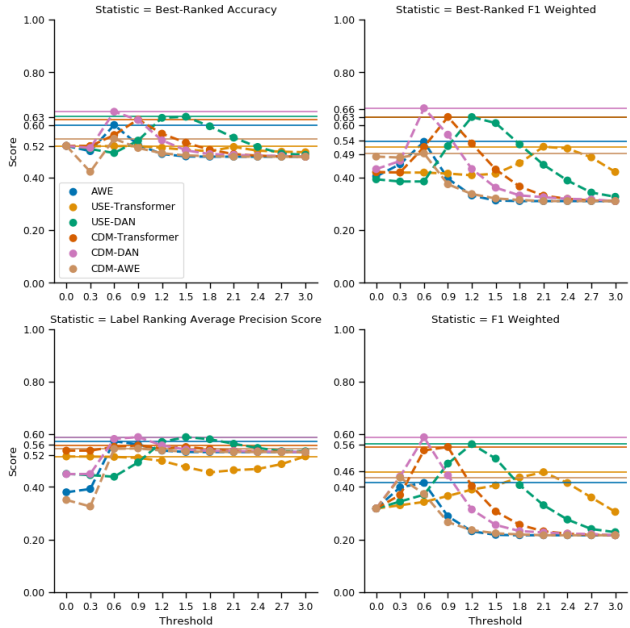

**Figure 5: Test-Set CB - Score comparison of the statistics of different algorithms when changing the threshold.**

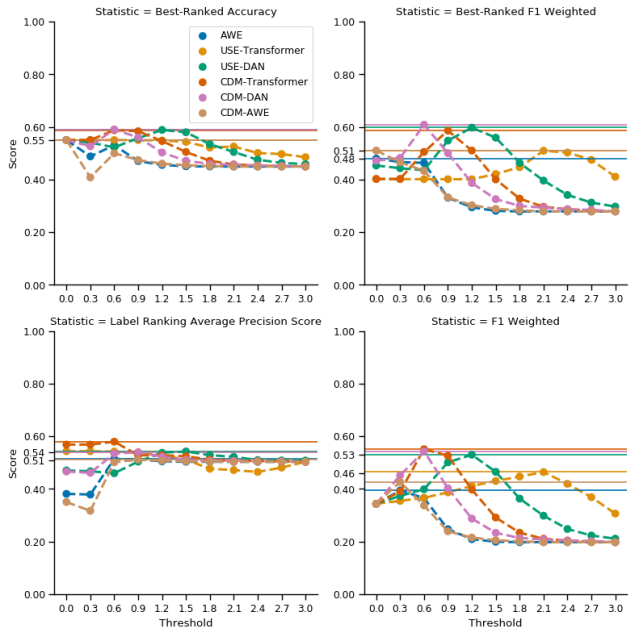

**Figure 6:Test-Set CL - Score comparison of the statistics of different algorithms when changing the threshold.**

We tested and compared 3 baselines against 2 variations of CDM, in order to understand which is the best:

- **CDM-DAN**: the algorithm described in Sections 4 and V.
- **CDM-Transformer**: it is like CDM-DAN, but with U computed through the Transformer-based USE.
- **AWE**: this algorithm is implemented like CDM but without the paradigmatic topic weighting, and with the TF-IDF cosine similarity F replaced by the GloVe AWE cosine similarity G. This is a baseline.
- **USE-DAN**: it is like AWE but with F replaced by the DAN-based USE cosine similarity (U). This is a baseline.
- **USE-Transformer**: it is like USE-DAN but with U computed through the Transformer-based USE. This algorithm performs much better than AWE. This is a baseline.

As we can see, CDM outperforms the baselines in both test-sets A and B, while in test-sets CB and CL it is less evident whether CDM is better than the baselines. It is worthy to remember that the annotation technique adopted for test-sets CB and CL increases the false negatives (that is the number of paragraphs labelled as not related to any SDG). The difference between the statistics achieved in test-sets CB and CL and those achieved in the other test-sets is quite evident. Anyway, the best algorithm seems to be CDM-DAN with threshold T = 0.6, both in the dev-set and in the other sets.

## 7. DISCUSSION

As we have seen, the technique we proposed achieves only modest accuracy performance on multi-label text classification. At this point a spontaneous question would be: is this enough?

We believe that improvements are possible, for sure, especially trying to exploit better transfer learning. Anyway our goal was not to produce the best algorithm possible for the task, nor to build an algorithm to replace humans in the task.

Our main goal was to produce a useful software for UN, that could help to speed up the process of qualifying the UN documents following the SDGs in order to monitor the progresses at the world level to fight poverty, discrimination, climate changes. In fact human labeling of UN documents would be a daunting task given the size of the impacted corpus. Thus, automatic labeling must be adopted at least as a first step of a multi-phase process to reduce the overall effort of cataloguing and classifying. Our algorithm can be used for this purpose, and it is going to be used by UN indeed.

Furthermore, one can argue that the hybrid approach we adopted can be obscure and unclear, because of the involved formula used to combine different similarity measures. These formula have been found using an empirical approach, without providing very strong theoretical guarantees of their efficacy, but we argue that they are much less obscure and un-explainable than any other deep learning algorithm usually adopted in state-of-the-art multi-label text classification.

Despite all the aforementioned issues, the technique we proposed is able to perform multi-label text classification of UN resolutions, introducing a less costly and time-consuming modeling solution. In fact, our approach combines TF-IDF and pre-trained SOTA DL models (such as the Universal Sentence Encoder) without any need for traditional transfer learning or any other expensive training procedure.

The adaptability of the proposed technique to different domains has not been thoroughly analyzed. Despite this, we believe that our technique can be adapted to work also in different domains, by:
1. Tuning the threshold T.
2. Carefully removing and/or adapting the instructions followed during the pre-processing phase (see Section 5.1).
3. Carefully choosing the documents used for building the TF-IDF model (see Section 5.2).
4. Changing the way the combined similarity C (see Section 5.3) is obtained, for example by giving more (or less) weight to the TF-IDF similarity, or by changing the paradigmatic topic weight.

## 8. RELATED WORK

As far as we know, most of the state of the art on AI for SDG has mainly focused on directly achieving SDGs by the use of AI [26,27,28], rather than identifying (as we did) whether SDGs have been indirectly (or not) mentioned inside texts. In this sense, our work is very different from previous ones, at least on this very specific topic. This is why we propose our work as baseline for further development, and we publish both the datasets and the source code used in all the experiments mentioned in this paper.[7]

On the other side, multi-label text classification is a well known problem and many solutions based on deep learning exist. Differently from [2], [21], our approach does not try to train a deep neural network for multi-label classification, but instead it tries to exploit together TF-IDF and pre-trained models like USE. Our way of combining TF-IDF with Average GloVe and USE differs from the one adopted in [22] or in [15], [16]. In fact, in [22], [15] the document embedding is obtained by weighting the word embeddings by their TF-IDF values, but in our approach we combine document similarities instead of word embeddings.[8]

## 9. CONCLUSIONS AND FUTURE WORK

We wanted to perform multi-label text classification of UN Resolutions through deep learning, but without any big-enough annotated training set. We designed a new ensemble method that effec-tively combines generic (non domain-specific) deep learning based document similarities with domain-specific TF-IDF document similarities, for achieving SDG classification of UN Resolutions.

The algorithm we described is quite versatile and powerful. In fact it is able to perform multi-label classification, it does not require much hyper-parameters tuning (practically only the value of the threshold T has to be tuned), it is super fast to train, it allows us to easily change the definitions of the classes without incurring significantly slower or more error-prone pre- processing and it performs quite well with relatively small training sets. The future work is to refine the gold standard with human experts and to apply the mark-up to the FAO Resolutions. Secondly we would like to use the same approach in order to classify the same UN Resolution texts with the UNBIS (United Nations Bibliographic Information System).

## ACKNOWLEDGMENTS

We would like to thank United Nations, in particular the High Level Committee on Management (HLCM), for organizing the Challenge UNGA Automatic Information Extraction and Knowledge Elicitation[9], and for their feedback and positive evaluation of the project[10], that won the challenge.

---

[7] https://github.com/Francesco-Sovrano/Deep-Learning-Based-Multi-Label-Text-Classification-of-UNGA-Resolutions

[8] In principle, the technique adopted in [22], [15] might be used to improve our algorithm.

[9] https://ideas.unite.un.org/unga-resolutions/Page/Home

[10] The project won the award of the UNGA Challenge: https://unite.un.org/news/university-bologna-team-wins-first-prize-united-nations-general-assembly-resolutions-extraction